\documentclass[fleqn,10pt]{SelfArx} 

\usepackage[english]{babel} 
\usepackage{bm}

\definecolor{color1}{RGB}{0,0,90} 
\definecolor{color2}{RGB}{0,20,20} 

\usepackage{hyperref} 

\hypersetup{
	hidelinks,
	colorlinks,
	breaklinks=true,
	urlcolor=color2,
	citecolor=color1,
	linkcolor=color1,
	bookmarksopen=false,
	pdftitle={Title},
	pdfauthor={Author},
}

\JournalInfo{March} 
\Archive{2022} 

\PaperTitle{Performance evaluation of volatility estimation methods for Exabel} 

\Authors{Øyvind Grotmol\textsuperscript{1}*, Martin Jullum\textsuperscript{2}, Kjersti Aas\textsuperscript{2}, Michael Scheuerer\textsuperscript{2}, ...} 
\affiliation{\textsuperscript{1}\textit{Exabel, Oslo, Norway}} 
\affiliation{\textsuperscript{2}\textit{Department for Statistical modeling, Machine learning and Artificial intelligence, Norsk Regnesentral, Oslo, Norway}} 
\affiliation{*\textbf{Corresponding author}: grotmol@exabel.com} 

\newcommand{\keywordname}{Keywords} 

\Abstract{
Quantifying both historic and future volatility is key in portfolio risk management. 
This note presents and compare estimation strategies for volatility estimation in an estimation universe consisting on 28 629 unique companies from February 2010 to April 2021, with 858 different portfolios. The estimation methods are compared in terms of how they rank the volatility of the different subsets of portfolios. 
The overall best performing approach estimates volatility from direct entitiy returns using a GARCH model for variance estimation.
}

\begin{document}

\maketitle 
\tableofcontents 
\thispagestyle{empty} 


\section{Introduction} 

The aim of this note is to describe and summarize a quantitative study comparing strategies for forecasting investment risk. 
The study concentrates on the investment risk of a portfolio relative to other portfolios, and uses volatility as the risk measure. 
Denote by $r_{k,t}$ the log-return of a company $k\in\mathbb{K}:=\{1,\ldots,K\}$ on day $t$ (hereafter for simplicity referred to as the return or company return). 
The (daily) volatility of company $k$ during a time period $\mathbb{T}$ of $|\mathbb{T}|$ trading days is then defined as the sample standard deviation
\begin{equation}
 \mathrm{vol}_k(\mathbb{T}) = \sqrt{\frac{1}{|\mathbb{T}|-1}\sum_{t\in\mathbb{T}}\left(r_{k,t}-\frac{1}{|\mathbb{T}|}\sum_{t\in\mathbb{T}} r_{k,t}\right)^2}, \label{eq:company-vol}
\end{equation}
with an analogous definition for portfolios. While it is common to scale $\mathrm{vol}_k(\mathbb{T})$ to reflect the volatility over $\mathbb{T}$ or a full year, we omit that here as we deem daily volatility more interpretable.

As a short-term investment tool, estimates of future volatility is more useful than computations of historic volatility. 
In this study, we will therefore consider methods for estimating (daily) volatility over consecutive test periods $\mathbb{T}$ of duration 3 months.
We will consider both direct approaches estimating the volatility using the historic observed returns of the companies, and indirect approaches using Exabel's factor model. 
We will also consider two different weighting strategies for data back in time: A naive window based approach which puts equal weight on all observations $q$ months back in time, and an approach which puts more weight on the most recent observations according to an estimated GARCH model.
Exabels's factor model is briefly described in Section \ref{sec:factor_model}, while all different volatility estimation methods are precisely defined in Section \ref{sec:vol}.

The estimation universe we will be working with here consists of a total of 28 629 companies from all over the world, with company information and returns ranging from February 2010 to April 2021.
As some companies have only existed for a portion of the full time period, the trading days are not the same for all companies. 
Due to this, and other missing data complications, some data pre-processing is required. 
We describe the removals, imputations and other actions we take to circumvent missing data issues in Section \ref{sec:dataprep}.

The volatilities are estimated for a range of portfolios, consisting of different companies from the estimation universe. The portfolios are defined based on the factors described in Section \ref{sec:factor_model} to reflect investment strategies following the different factors. In total we are estimating volatilities for 858 different portfolios over 97 different 3-month test periods $\mathbb{T}$. See Section \ref{sec:pf} for details.

The estimation methods will be compared and measured according to their ability to rank portfolios based on their risk. 
We will do this by computing Kendall's $\tau$ \cite{kendall1938new} between the estimated volatilities and the target volatilities of a set of portfolios (after the latter have been observed). 
Details about the performance evaluation is provided in Section \ref{sec:perf}.
Different sets of portfolios are considered, and their results are presented in Section \ref{sec:res}.
Some conclusions are provided in Section \ref{sec:summary}.


\section{Factor model}
\label{sec:factor_model}

With a factor model we would like to explain the returns of all companies in $\mathbb{K}$ as well as possible through a small number of interpretable factors, i.e.\ attributes of a company which appear to influence its return. 
The {\em factor loading} $X_{k,t}^*$ of company $k$ quantifies its exposure to a particular factor $*$ at day $t$. 
Three different types of factor loadings are considered in Exabel's factor model:
\begin{itemize}[noitemsep] 
    \item {\em Style}-loadings $X_{k,t}^s, \; s\in\mathbb{S}$ (set of style factors)
    \item {\em Country}-loadings $X_{k,t}^c, \; c\in\mathbb{C}$ (set of countries)
    \item {\em Industry}-loadings $X_{k,t}^i, \; i\in\mathbb{I}$ (set of industries)
\end{itemize}
Using a cross-sectional regression model, we then want to estimate (separately for each day) {\em factor returns} $f^{s}_t, s\in\mathbb{S}, f^c_t, c\in\mathbb{C}$, and $f^i_t, i\in\mathbb{I}$ together with the overall market return $f^m_t$:
\begin{equation}
 r_{k,t} = f^m_t + \sum_{s\in\mathbb{S}} X_{k,t}^s f^s_t + \sum_{c\in\mathbb{C}} X_{k,t}^c f^c_t + \sum_{i\in\mathbb{I}} X_{k,t}^i f^i_t + \epsilon_{k,t}, \label{eq:factor-model}
\end{equation}
where $\epsilon_{k,t}\sim\mathcal{N}(0,\sigma_{k,t}^2)$ is the residual term that accounts for the fraction of the return $r_{k,t}$ not explained by the company's exposure to the different factors. 
In Exabel's factor model, the factor loadings $X_{k,t}^{*}$ are updated on a monthly basis, i.e.~they are constant within every month.
All style factors $X_{k,t}^{s}$ are market cap centered, meaning that they satisfy $\sum_{k \in \mathbb{K}} mc_{k,t}X_{k,t}^s$, where  $mc_{k,t}$ denotes the market capitalization of company $k$.
Furthermore, the estimates $\widehat{f}^*_t$ for the factor $f^*_t$ are obtained using a constrained weighted least squares approach where the companies are weighted by $mc_{k,t}$. See \cite{Grotmol-exabels-factor-model} for further details about Exabel's factor model.

In total there are $l = 77$ different factors. Using matrix notation, \eqref{eq:factor-model} can be written as
\begin{align}
    r_{k,t} = \mathbf{X}_{k,t}^\top \mathbf{f}_t + \epsilon_{k,t}, \label{eq:matrix_factormodel}
\end{align}
where  $\mathbf{f}_t$ is the $l$-dimensional vector of all factor returns, and $\mathbf{X}_{k,t}$ is the $l$-dimensional vector of all factor loadings (we define $X^m_{k,t}=1$) .


\section{Volatility estimation and computation}
\label{sec:vol}

As mentioned, the purpose of this note is to compare different methods for estimating the future volatility of portfolios. In this section we extend the single company volatility definition in \eqref{eq:company-vol} to the volatility of a portfolio, and describe the different methods we will be using for estimating these ahead in time. 
All quantities in this section are specific to a portfolio $\mathbb{P}=\{k_1,\ldots,k_p\}$ of $p$ companies, with corresponding portfolio weights $w_{k,t}$ for company $k$ at day $t$. Every portfolio has the property that $\sum_{k=1}^p |w_{k,t}|=1$ for all $t$.
To simplify notation, we will, however, often omit $\mathbb{P}$. When it is clear from the context, we will also sometimes omit the test period $\mathbb{T}$ from our notation.

\subsection{Target portfolio volatility}
\label{sec:target_vol}

We start by defining the target portfolio volatility, i.e.~the future volatility that the different estimation methods will attempt to predict. 
As mentioned in the introduction, we will consider consecutive 3-months test periods $\mathbb{T}$, and shall for simplicity assume that all portfolios have fixed portfolio weights $w_{k,t}$ within each of the test periods $\mathbb{T}$, i.e. $w_{k,t}=w_k$ for all $t \in \mathbb{T}, k \in \mathbb{P}$. With these simplifications in place, we define the target portfolio volatility for time period $\mathbb{T}$ as
\begin{align}
    \text{vol}(\mathbb{T}) = \sqrt{\mathbf{w}^\top {\Sigma}(\mathbb{T}) \mathbf{w}}, \label{eq:vol_portfolio}
\end{align}
where $\mathbf{w} = (w_{1},\ldots,w_{p})^\top$ is the vector of portfolio weights in time period $\mathbb{T}$.
Furthermore, ${\Sigma}(\mathbb{T})$ is the $p \times p$-dimensional sample covariance matrix of the daily returns in time period $\mathbb{T}$, for the companies in $\mathbb{P}$.

An alternative to the portfolio volatility definition in \eqref{eq:vol_portfolio} would be to compute the daily returns of the portfolio first, and then compute the variance of these returns across $\mathbb{T}$. This is, however, deemed more sensitive to missing company returns, as missing a single return for one day with nonzero weight would leave the portfolio return for that day undefined. See Section \ref{sec:dataprep} for how we handle missing company returns when estimating the covariance matrix ${\Sigma}(\mathbb{T})$.

\subsection{Estimating future portfolio volatility}

In this section we discuss methods to estimate the target portfolio volatility in Section \ref{sec:target_vol} using historical data.
We will discuss two types of approaches. 
The first uses the actual returns directly, while the second is an indirect approach which uses the factor model discussed in Section \ref{sec:factor_model}.
Both approaches uses the portfolio volatility formula in \eqref{eq:vol_portfolio} to estimate the future portfolio volatility, 
except that the (future) sample covariance matrix ${\Sigma}(\mathbb{T})$ (which is unknown at the estimation time point) is replaced with an estimate.
 In the subsequent subsections, we first present general formulae for the two types of estimating approaches, before we provide precise estimation methods for the unknown quantities.

\subsubsection{The direct return approach}
\label{sec:direct}

For the direct (company) return approach, the estimate of the future covariance matrix takes the form
\[ \widehat{\Sigma}_{\text{r}} = \widehat{D}_{\text{r}}  \widehat{R}_{\text{r}}  \widehat{D}_{\text{r}}, \]
where $\widehat{R}_{\text{r}}$ is an estimate of the $p \times p$ dimensional future correlation matrix of the returns, and $\widehat{D}_{\text{r}}$ is a $p \times p$ diagonal matrix with estimates of the standard deviation of the company returns on the diagonal. The portfolio volatility estimate using the direct return approach (r) thus takes the form
\begin{align}
        \widehat{\text{vol}}_{\text{r}}(\mathbb{T}) = \sqrt{\mathbf{w}^\top \widehat{\Sigma}_{\text{r}} \mathbf{w}}. \label{eq:vol_portfolio_direct}
\end{align}

\subsubsection{The indirect factor based approach}
\label{sec:indirect}

To write up a general volatility expression for the indirect factor based approach (f), recall the matrix formulation of the estimated factor model in \eqref{eq:matrix_factormodel}. With estimated parameters, this factor model takes the form $\widehat{r}_{k,t} = \mathbf{X}_{k,t}^\top \widehat{\mathbf{f}}_t$, where $\widehat{\mathbf{f}}_t$ is the estimated analogue of $\mathbf{f}_t$. Let us also denote by $L$ the $l \times p$ dimensional matrix obtained by columnwise concatenation of the $l$-dimensional vectors $\mathbf{X}_{1,t}, \mathbf{X}_{2,t}, \ldots, \mathbf{X}_{p,t}$.
For the indirect, factor based approach, the estimate of the future covariance may then be written as 
\begin{align}
\widehat{\Sigma}_{\text{f}} =  L^\top  \widehat{\Sigma}_{\text{f}}^{\text{f}}  L = 
L^\top \widehat{D}_{\text{f}}  \widehat{R}_{\text{f}}  \widehat{D}_{\text{f}} L, \label{eq:Sigma_f}
\end{align}
where $\widehat{\Sigma}_{\text{f}}^{\text{f}}$ and $\widehat{R}_{\text{f}}$ are estimates of the $l \times l$ dimensional future covariance and correlation matrices of the factors $\widehat{\mathbf{f}}_t$. 
The $\widehat{D}_{\text{f}}$ is an $l \times l$ diagonal matrix with estimates of the (future) standard deviation of the factors at the diagonal. 

When using the factor model to estimate the volatility, we also need to account for the idiosyncratic variance $\sigma^2_{k,t}$ of the residual term $\epsilon_{k,t}$. 
As the factor model assumes that the error terms are independent, the estimated portfolio variance then becomes $\sum_{k \in \mathbb{P}} w^2_{k,t}\widehat{\sigma}^2_{k,t}$, which in matrix form may be written $\mathbf{w}^\top \widehat{\boldsymbol{\sigma}}^2_{\mathbb{P}}\mathbf{w}$, where 
$\widehat{\boldsymbol{\sigma}}^2_{\mathbb{P}}$ is the $p \times p$-dimensional diagonal matrix with estimates for the variances of the companies in the portfolio $\mathbb{P}$ on the diagonal.
Combining the insertion of $\widehat{\Sigma}_{\text{f}}^{\text{f}}$ for ${\Sigma}(\mathbb{T})$ in \eqref{eq:vol_portfolio} with the additional residual variance term, we get the following general formula for the portfolio volatility estimate using the factor based approach (f): 
\begin{align}
        \widehat{\text{vol}}_{\text{f}}(\mathbb{T}) &=  \sqrt{\mathbf{w}^{\top} \widehat{\Sigma}_{\text{f}}  \mathbf{w} + \mathbf{w}^\top \widehat{\boldsymbol{\sigma}}^2_{\mathbb{P}}\mathbf{w}} . \label{eq:vol_portfolio_factor0}
\end{align}
Furthermore, by pre-computing a $l \times p$-dimensional portfolio factor loading matrix $L_{\mathbb{P}} = \mathbf{w}^{\top}L^\top$, we don't need to compute the full $p\times p$ dimensional $\widehat{\Sigma}_{\text{f}}$, but only the $l\times l$-dimensional $\widehat{\Sigma}_{\text{f}}^{\text{f}}$. Thus, \eqref{eq:vol_portfolio_factor0} can also be written as 
\begin{align}
        \widehat{\text{vol}}_{\text{f}}(\mathbb{T}) &= \sqrt{L_{\mathbb{P}}^{\top} \widehat{\Sigma}_{\text{f}}^{\text{f}}  L_{\mathbb{P}} + \mathbf{w}^\top \widehat{\boldsymbol{\sigma}}^2_{\mathbb{P}}\mathbf{w}}, \label{eq:vol_portfolio_factor}
\end{align}
which is computationally more efficient for portfolios with many companies.

\subsubsection{Estimating the unknown quantities}

The formulae for the estimated volatility in \eqref{eq:vol_portfolio_direct} and \eqref{eq:vol_portfolio_factor} requires specification of the estimated covariance matrices
$\widehat{\Sigma}_{\text{r}} = \widehat{D}_{\text{r}}  \widehat{R}_{\text{r}}  \widehat{D}_{\text{r}}$ and 
$\widehat{\Sigma}_{\text{f}}^{\text{f}} = \widehat{D}_{\text{f}}  \widehat{R}_{\text{f}}  \widehat{D}_{\text{f}}$, respectively. 
We will provide two different methods for estimating each of them based on historic data, in addition to an estimation method for the residual variances in the factor model.

\subsubsection*{Naive window estimation method}
The simplest way to estimate a covariance matrix for future companies/factors is to use the sample covariance matrix for data in a window covering a certain number of days back in time. This is a naive estimation method which weights all company/factor returns within the window equally. If the window is too long one risks that the oldest company/factor returns are not representative, while a too short window risks giving unstable estimates as there is not enough data. 
Let us denote by $\widehat{\Sigma}_{\text{r,naive}}(q)$ and $\widehat{\Sigma}_{\text{f,naive}}^{\text{f}}(q)$ the company/factor return sample covariance matrices with a window covering $q$ months prior to the test period $\mathbb{T}$.

\subsubsection*{GARCH model for variance estimation}
The naive method above has the problem of requiring a rather long period (large $q$) in order to get stable and good estimates, while at the same time such a long period makes it react rather slow to changes in the volatility over time. One way to overcome this is to put more weight on the more recent observations than the older ones. Our approach in that direction is to modify the estimation of the covariance matrices by combining the naive window method for the correlation structure $\widehat{R}_{\text{r}}$ and $\widehat{R}_{\text{f}}$ with separate univariate GARCH models \cite{bollerslev1986generalized} for the variances of the companies ($\widehat{D}_{\text{r}}$ and $\widehat{D}_{\text{f}}$) . There exists multivariate GARCH models that could estimate covariance structure directly, but in the high dimensional settings present here, that would be computationally infeasible. GARCH stands for generalized autoregressive conditional heteroskedasticity, and is essentially an autoregressive moving average (ARMA) model on the error variance of a time series model. Here we will just assume a constant mean model, and use both autoregressive (AR) and moving average (MA) terms of order 1 for the variance. This is the most commonly used GARCH form within financial modelling \cite{bollerslev1986generalized}. For company $k$, this GARCH model takes the form
\begin{align}
    r_{k,t} &= \mu_k + \varepsilon_{G,k,t}, \notag \\
    \varepsilon_{G,k,t}|\varepsilon_{G,k,t-1},\sigma^2_{G,k,t-1} &\sim N(0,\sigma^2_{G,k,t}), \notag \\
    \sigma^2_{G,k,t} &= \omega_k + \alpha_k\varepsilon^2_{G,k,t-1} + \beta_k \sigma^2_{G,k,t-1}, \notag
\end{align}
where $\mu_k, \omega_k, \alpha_k$ and $\beta_k$ are parameters that have to be estimated. For the factor model, the formulation is completely analogous, except that company returns $r_{k,t}$ are replaced by the fitted factor returns $\widehat{f}^*_t$.
When estimating the parameters in these model specifications, we will always use a history of 3 years prior to the test period $\mathbb{T}$. 
The diagonal matrices $\widehat{D}_{\text{r}} = \widehat{D}_{\text{r,GARCH}}$ and $\widehat{D}_{\text{f}} = \widehat{D}_{\text{f,GARCH}}$ are then filled with the modelled company/factor return variances for the latest observed day. 
Combining these diagonal variance matrices with naive window based estimators for the correlation structure (i.e.~$\widehat{R}_{\text{r,naive}}(q)$ and $\widehat{R}_{\text{f,naive}}(q)$), we get
\begin{align}
\widehat{\Sigma}_{\text{r,GARCH}}^{\text{f}}(q) = \widehat{D}_{\text{r,GARCH}}  \widehat{R}_{\text{r,naive}}(q)  \widehat{D}_{\text{r,GARCH}}, \notag \\
\widehat{\Sigma}_{\text{f,GARCH}}^{\text{f}}(q) = \widehat{D}_{\text{f,GARCH}}  \widehat{R}_{\text{f,naive}}(q)  \widehat{D}_{\text{f,GARCH}}. \notag
\end{align}

\subsubsection*{Factor model residual variance}
Regardless of whether the naive window method or the GARCH model is used to estimate $\widehat{\Sigma}_{\text{f}}^{\text{f}}$ for the factor models, we use the basic sample variance of the factor residuals $\epsilon_{k,t}$ for $t$ in a window $q$ months prior to $\mathbb{T}$ as an estimate of $\sigma^2_{k,t}, k \in \mathbb{P}$. We use the same window length $q$ as for $\widehat{\Sigma}_{\text{f,naive}}^{\text{f}}(q)$ and $\widehat{\Sigma}_{\text{f,GARCH}}^{\text{f}}(q)$, and denote the estimate of the residual variance diagonal matrix $\boldsymbol{\sigma}^2_{\mathbb{P}}$ by $\widehat{\boldsymbol{\sigma}}^2_{\mathbb{P}}=\widehat{\boldsymbol{\sigma}}^2_{\mathbb{P},\text{naive}}(q)$. 

\subsubsection{Final estimation schemes}
\label{sec:finalest}
To summarize, we have the following types of estimation schemes for the portfolio variance
\begin{align}
\widehat{\text{vol}}_{\text{r},\text{GARCH}}(q) &= \sqrt{\mathbf{w}^\top \widehat{\Sigma}_{\text{r,GARCH}}(q) \mathbf{w}}, \notag \\
\widehat{\text{vol}}_{\text{r},\text{naive}}(q) &= \sqrt{\mathbf{w}^\top \widehat{\Sigma}_{\text{r,naive}}(q) \mathbf{w}}, \notag \\
\widehat{\text{vol}}_{\text{f},\text{GARCH}}(q) &= \sqrt{L_{\mathbb{P}}^{\top} \widehat{\Sigma}_{\text{f,GARCH}}^{\text{f}}(q)  L_{\mathbb{P}} + \mathbf{w}^\top \widehat{\boldsymbol{\sigma}}^2_{\mathbb{P},\text{naive}}(q)\mathbf{w}}, \notag \\
\widehat{\text{vol}}_{\text{f},\text{naive}}(q) &= \sqrt{L_{\mathbb{P}}^{\top} \widehat{\Sigma}_{\text{f,naive}}^{\text{f}}(q)  L_{\mathbb{P}} + \mathbf{w}^\top \widehat{\boldsymbol{\sigma}}^2_{\mathbb{P},\text{naive}}(q)\mathbf{w}} \notag, 
\end{align}
which will be combined with different window lengths $q$.

\section{Data pre-processing}
\label{sec:dataprep}

In this section we briefly describe the data in our estimation universe, and list the data processing steps which filter or adjusts the original data.

Our data set consists of daily returns and monthly factor loadings for a total of 28 629 unique companies from February 2010 to April 2021. From these, we have estimated factor returns corresponding to Exabel's factor model, described in Section \ref{sec:factor_model}, with further details in \cite{Grotmol-exabels-factor-model}.

We have made the following adjustments and filtering of our data:
\begin{itemize}
    \item Missing industry and country loadings $X^c_{k,t}$ and $X^i_{k,t}$ are filled with 0.
    \item Missing factor returns $f^*_{k,t}$ are filled with 0.
    \item Missing market caps are replaced by the square of the factor loading company weights.
    \item Companies with no registered factor loadings for a certain month are excluded from the study that entire month.
    \item Companies with no registered company return a certain month are excluded from the study that entire month.
    \item All sample covariance/correlation matrices are estimated pairwise using daily data, after removing all missing pairs. Non-estimable variances or covariances (due to completely missing data) are replaced by 0.
    \item We use the naive sample variance instead of the GARCH estimate for single companies/factors when there is either less than 100 observed company/factor returns, the GARCH model does not converge, or it is not able to produce an estimate for other reasons.  
    \item When estimating the covariance/correlation matrices pairwise, and adjusting non-estimable elements one may end up with covariance/correlation matrices which are not (computationally) positive definite. In such cases, we replace it by the nearest positive definite matrix using the algorithm of \cite{higham2002computing}.
\end{itemize}

As a consequence of the above filtration and adjustments, companies that have missing observations in the entire estimation window ($q$ months) contribute with zero volatility to the portfolios they are involved in.



\section{Portfolios}
\label{sec:pf}

This section describes and summarizes the portfolios we use in our study. 
The portfolios are constructed based on the factors described in Section \ref{sec:factor_model} and the market cap $mc$, to reflect different investment strategies. 
We construct a total of 286 different portfolios for the entire time span. 
These consist of both unrestricted, region restricted and subregion restricted factor based portfolios. 
We consider the following regions: America, Asia and Europe, and the following subregions: 
Latin America and the Caribbean, Northern America, Eastern Asia, South Eastern Asia, Southern Asia, Western Asia, Northern Europe, Southern Europe, and Western Europe. The regions are defined using the ISO 3166 country codes and region/subregion definitions.
Each portfolio has a minimum of 40, and a maximum of 300 companies in each test period $\mathbb{T}$. 
The portfolio weights $w_{k,t}$ for company $k$ and $t \in \mathbb{T}$ are based on the factor loadings the last month before $\mathbb{T}$.
In the portfolios, long positions have positive weights, while short positions have negative weights. 
We will both consider portfolios with purely long positions, and market neutral portfolios where $\sum_{k=1}^p w_{k,t} = 0$. 
All portfolios are scaled such that $\sum_{k=1}^p |w_{k,t}| = 1$.
We only allow companies which have a market cap of at least \$200 million. 
All long/short portfolios are required to have at least 20 companies with positive weight and 20 companies with negative weight.

When constructing the portfolios below, we will be filtering and weighting companies based on both factor loadings and the market cap $mc_{k,t}$. Although $mc$ has no factor loading, we will use that term also there to simplify the description. Thus, when construction portfolios weighted by $mc$ below, the term `factor loading' and the notation $X^*_{k,t}$ will refer to $mc_{k,t}$.
All portfolio weights are constructed using factor loadings $X^*_{k,t}$ at the last time point $t$ prior to each test period $\mathbb{T}$. 
Our portfolios are constructed as follows:
\begin{itemize}
    \item {\bf Long, unrestricted portfolios, all factors + $mc$:} For each test period $\mathbb{T}$ and each of the 77 factors + $mc$, we filter out all companies without a \textit{positive} factor loading $X^*_{k,t}$. 
    Out of these, we keep the (maximum) 300 companies with the largest market cap. 
    For the remaining companies, the portfolio weight $w_{k,t}$ is set proportional to $X^*_{k,t}$.
    \item {\bf Long/short, unrestricted portfolios, all style factors:} For each test period $\mathbb{T}$ and each of the $|\mathbb{S}|=11$ style factors, we filter out all companies without a \textit{nonzero} factor loading $X^*_{k,t}$. 
    Out of these, we keep the (maximum) 300 companies with the largest market cap. 
    For the remaining companies, the portfolio weight $w_{k,t}$ is set proportional to $X^s_{k,t}$.
    \item {\bf Long, region/subregion portfolios, style factors + $mc$:} For each test period $\mathbb{T}$, each region/subregion and each of the $|\mathbb{S}|=11$ style factors + $mc$, we filter out all companies without a positive country factor loading  $X^c_{k,t}$ for at least one of the countries in the relevant region/subregion. 
    We then filter out all companies without a \textit{positive} factor loading $X^*_{k,t}$.
    Out of these, we keep the (maximum) 300 companies with the largest market cap. 
    For the remaining companies, the portfolio weight $w_{k,t}$ is set proportional to $X^*_{k,t}$.
    \item {\bf Long/short, region/subregion portfolios, style factors:} For each test period $\mathbb{T}$, each region/subregion and each of the $|\mathbb{S}|=11$ style factors, we filter out all companies without a positive country factor loading  $X^c_{k,t}$ for at least one of the countries in the relevant region/subregion.
    When then filter out all companies without a \textit{nonzero} factor loading $X^*_{k,t}$.
    Out of these, we keep the (maximum) 300 companies with the largest market cap. 
    For the remaining companies, the portfolio weight $w_{k,t}$ is set proportional to $X^s_{k,t}$.
\end{itemize}

\begin{figure}
    \centering
    \includegraphics[width=\linewidth]{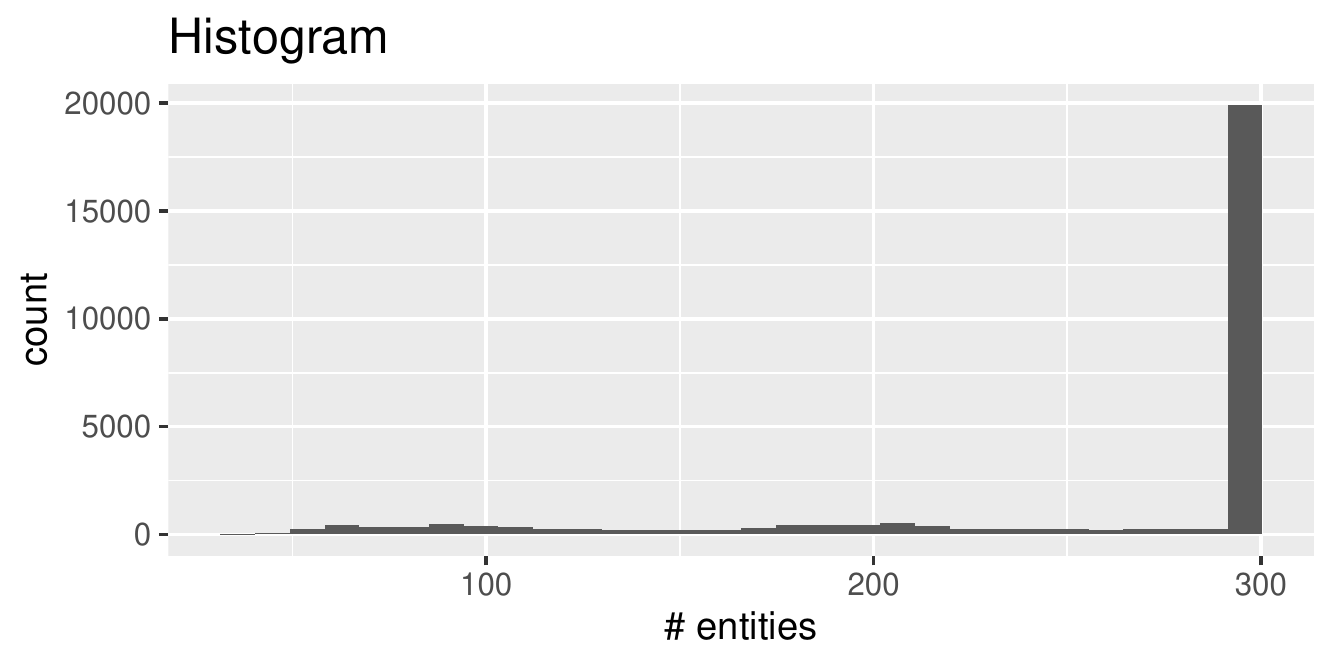}
    \caption{Histogram with number of companies per test period $\mathbb{T}$ and original portfolio.}
    \label{fig:histogram1}
\end{figure}

The number of companies per test period $\mathbb{T}$ vary between 40 and 300 as shown in the histogram in Figure \ref{fig:histogram1}.
Table \ref{tab:portfolios} shows the number of portfolios of each type.
\begin{table}[ht!]
    \centering
    \begin{tabular}{lllr}
      \toprule
        Long/short & Restriction & Type & \#  \\
         \midrule
        Long & Unrestricted & country & 27\\
        Long & Unrestricted & industry & 12\\
        Long & Unrestricted & style $+mc$ & 12\\
        Long/short & Unrestricted & style & 11\\
        Long & 3 regions & style $+mc$ & 36\\
        Long/short & 3 regions & style & 33\\
        Long & 9 subregions & style $+mc$ & 79\\
        Long/short & 9 subregions & style & 76\\
\bottomrule        
    \end{tabular}
    \caption{The number of original portfolios of each type.}
    \label{tab:portfolios}
\end{table}

For each of the 286 portfolios above, we generate two additional `random' portfolios. Each of these are constructed by sampling 50 companies with replacement from the original portfolio, using $w_{k,t}$ as sampling weights. 
This gives 584 additional portfolios, such that we end up with a total of 876 unique portfolios. 
The purpose of including the additional `random' portfolios is to control the sensitivity to strictly factor based portfolio definitions.

\begin{figure}
    \centering
    \includegraphics[width=\linewidth]{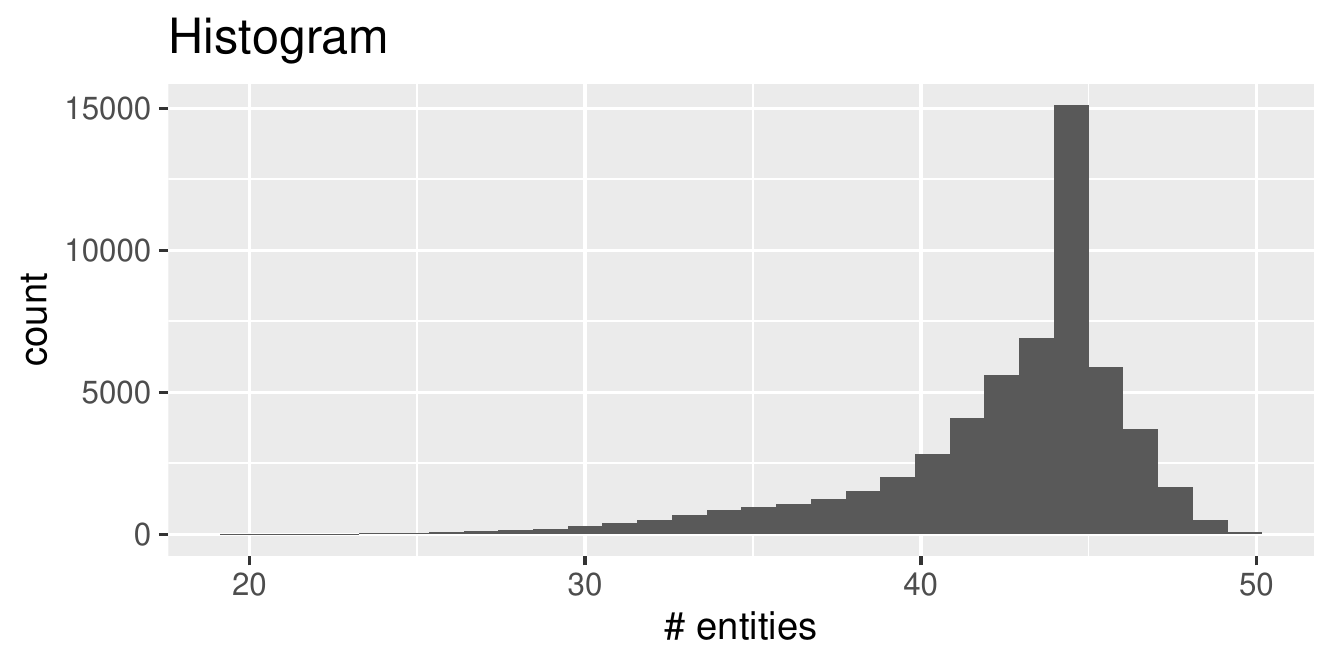}
    \caption{Histogram with number of companies per test period $\mathbb{T}$ and random portfolio.}
    \label{fig:histogram2}
\end{figure}

For the random portfolios, the number of companies per test period $\mathbb{T}$ vary between 18 and 50 as shown in the histogram in Figure \ref{fig:histogram2}.

\section{Performance evaluation}
\label{sec:perf}
\begin{figure*}[ht]\centering
    \includegraphics[width=0.8\linewidth]{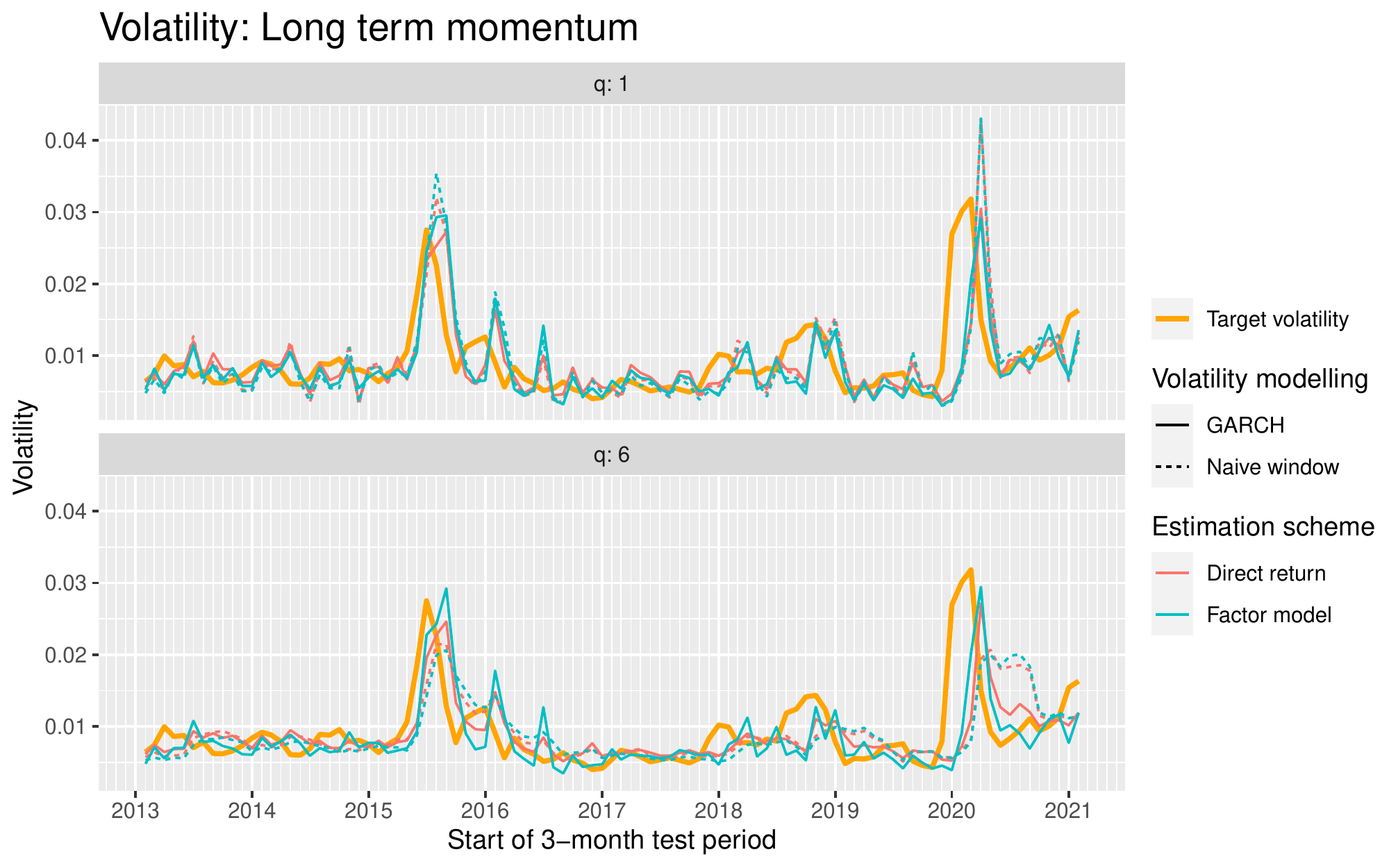}
    \caption{Estimated volatility compared to target volatility across test periods for a portfolio based on long term momentum. The upper and lower panels shows, respectively, the volatilities corresponding to the $q=1$ and $q=6$ month estimation windows.}
    \label{fig:time_plot}
\end{figure*}

In this section we describe how we evaluate the performance of the volatility estimation methods described in Section \ref{sec:vol}. 
When comparing different volatility estimation methods, we are mainly interested in their ability to correctly rank the portfolios according to their volatility (i.e.~their investment risk). 
As a consequence, we will evaluate the performance of the estimation methods by computing Kendall's $\tau$ \cite{kendall1938new} between the estimated volatilities and the target volatilities for sets of portfolios within each test period $\mathbb{T}$. 
Kendall't $\tau$ measures the ordinal association between the estimated and target volatilities by comparing their sorting order, and thereby ignoring the actual numeric values. 
It takes the value 1 if the ordering is exactly the same, and 0 if there is no association between their orderings. 
We use Kendall's $\tau$-b, which also accounts for ties in the ordering, see e.g.~\cite{agresti2010analysis}.
Since long and long/short portfolios are fundamentally different, we will compare these portfolio sets separately. 
We will evaluate the rankings of the full set of portfolios (still long and long/short separately), but also various subsets of portfolios. 
In particular we will perform separate computations for the original and random portfolios to check to what extent the results generalize.
The performance across all test periods $\mathbb{T}$ will be summarized by taking the mean of the Kendall's $\tau$'s over all test periods.

\section{Results}
\label{sec:res}

In this section we summarize the results from the performance study. We compare the direct return and factor model approaches, using both the naive window and GARCH based estimators. 
This gives us the 4 types of approaches listed in Section \ref{sec:finalest}, each combined with estimation window lengths $q=1, 3, 6$ and $12$ months. 
This gives us a total of 16 different estimation methods. 
These will be compared across a total of 97 test periods $\mathbb{T}$ of duration 3 months. 
The first test period $\mathbb{T}$ covers February 2013 to April 2013, the second covers March 2013 to May 2013, and so on, until the 97th test period which covers February 2021 to April 2021.
Although we focus on the ordering ability of the different methods, we first provide a figure showing how well the different methods estimate the target volatility for an unrestricted long portfolio weighted based on the long term momentum style factor, see Figure \ref{fig:time_plot}.
When investigating this plot, it is important to be aware two concepts  1) The x-axis shows the start of the 3 month period, i.e. the 2020 mark indicates the volatility during the period Jan 2020 throughout March 2020. 2) Volatilities are computed over consecutive 3-month periods, meaning that increased volatility in a single month will be spread out on three time periods.
From Figure  \ref{fig:time_plot} we see see that the estimation schemes has a natural delay in the reaction to changes in the target volatility. 
For the naive window method, we see as expected that the curve for $q=6$ months is less erratic, and holds onto a higher volatility level longer than the curve $q=1$ month.
While the GARCH and naive window methods are quite similar for this portfolio, the GARCH model reacts faster and does not seem to overcompensate, see in particular the late reaction to reduced volatility for $q=6$ in the middle of 2020, and the overcompensation for $q=1$ early in 2020. 
The factor model and the direct return estimation schemes are also quite similar for this portfolio, perhaps with a mild tendency towards less erratic behavior for the direct return approach, at least for $q=6$.
Note that the picture may look completely different for other portfolios. 

\begin{figure*}[ht]\centering
    \includegraphics[width=0.8\linewidth]{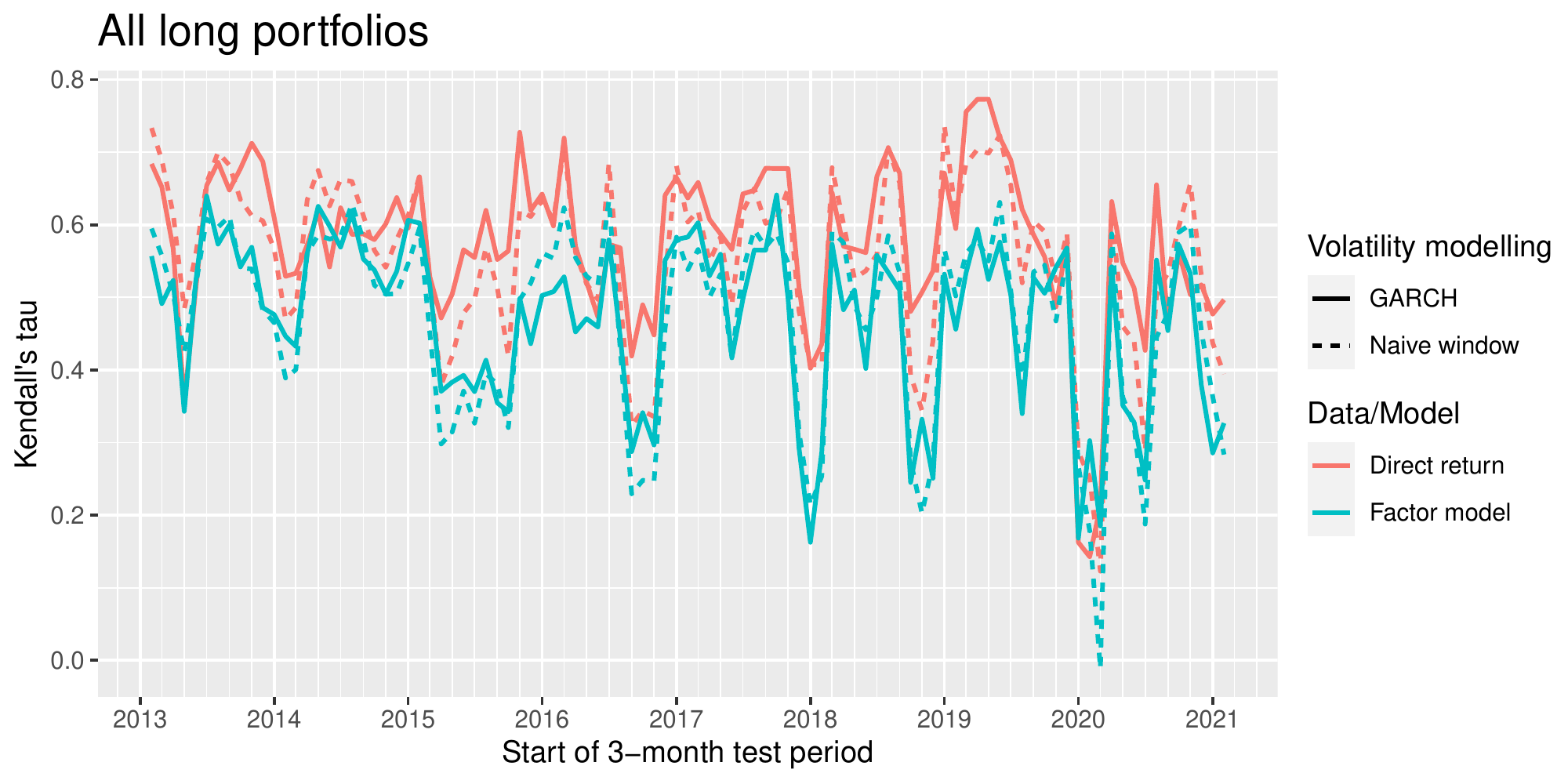}
    \caption{Kendall's $\tau$ between target and estimated volatility across test periods for all the 498 long portfolios. Only the results for the best performing window lengths are shown for each type of approach.}
    \label{fig:All.long}
\end{figure*}
\begin{table}[ht]
\centering
\begin{tabular}{lllr}
  \toprule
Data/Model & Volatility modelling & q & avg($\tau$) \\ 
  \midrule
Direct return & GARCH & 12 & 0.576 \\ 
  Direct return & GARCH & 6 & 0.570 \\ 
  Direct return & GARCH & 3 & 0.558 \\ 
  Direct return & Naive window & 6 & 0.554 \\ 
  Direct return & Naive window & 3 & 0.539 \\ 
  Direct return & Naive window & 12 & 0.539 \\ 
  Direct return & GARCH & 1 & 0.519 \\ 
  Direct return & Naive window & 1 & 0.492 \\ 
  Factor model & GARCH & 6 & 0.470 \\ 
  Factor model & Naive window & 6 & 0.470 \\ 
  Factor model & GARCH & 12 & 0.464 \\ 
  Factor model & GARCH & 3 & 0.463 \\ 
  Factor model & Naive window & 12 & 0.458 \\ 
  Factor model & Naive window & 3 & 0.456 \\ 
  Factor model & GARCH & 1 & 0.434 \\ 
  Factor model & Naive window & 1 & 0.422 \\ 
   \bottomrule
\end{tabular}
\caption{Average Kendall tau for portfolio group 'All long'. q denotes the window length in months for the correlation structure (and also the variance structure for the naive window method)} 
\label{tab:All.long}
\end{table}

Turning to the ordering ability of the different methods, we first consider the ability to order all the 498 different long portfolios (both original and random portfolios, which are either unrestricted or restricted to certain regions/subregions). 
Figure \ref{fig:All.long} shows Kendall's $\tau$ across all test periods.  
Only the best performing window length $q$ (measured by avg$(\tau)$) is plotted for each of the 4 approaches.
We see that except for a few test periods, such as those around the turn of the year 2019/2020, the direct return approach performs better than the factor model. For both estimation schemes, the GARCH volatility model almost always performs better than the naive window approach. 
Table \ref{tab:All.long} shows the average of the Kendall's $\tau$ across all the test periods $\mathbb{T}$.
It shows that overall, all direct approaches perform better than the factor models. 
\begin{table}[ht]
\centering
\begin{tabular}{lllr}
  \toprule
Data/Model & Volatility modelling & q & avg($\tau$) \\ 
  \midrule
Direct return & GARCH & 6 & 0.645 \\ 
  Direct return & GARCH & 12 & 0.645 \\ 
  Direct return & Naive window & 6 & 0.639 \\ 
  Direct return & GARCH & 3 & 0.629 \\ 
  Direct return & Naive window & 12 & 0.624 \\ 
  Direct return & Naive window & 3 & 0.620 \\ 
  Direct return & GARCH & 1 & 0.587 \\ 
  Direct return & Naive window & 1 & 0.573 \\ 
  Factor model & Naive window & 6 & 0.570 \\ 
  Factor model & Naive window & 3 & 0.569 \\ 
  Factor model & GARCH & 6 & 0.569 \\ 
  Factor model & GARCH & 3 & 0.566 \\ 
  Factor model & GARCH & 12 & 0.560 \\ 
  Factor model & Naive window & 12 & 0.556 \\ 
  Factor model & GARCH & 1 & 0.552 \\ 
  Factor model & Naive window & 1 & 0.549 \\ 
   \bottomrule
\end{tabular}
\caption{Average Kendall tau for portfolio group 'All long/short'. q denotes the window length in months for the correlation structure (and also the variance structure for the naive window method)} 
\label{tab:All.long.short}
\end{table}

\begin{figure*}[!htbp]\centering
    \includegraphics[width=0.8\linewidth]{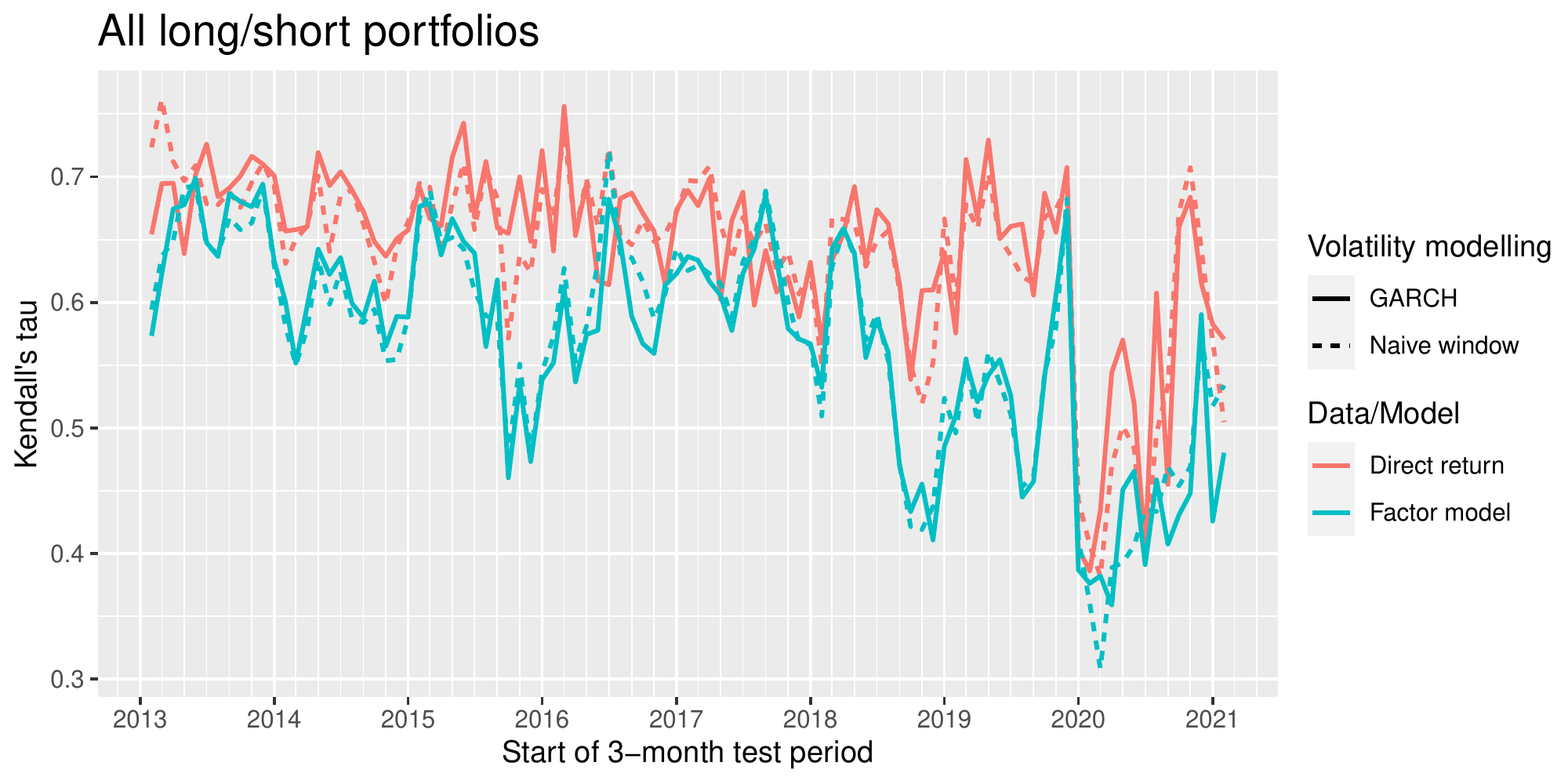}
    \caption{Kendall's $\tau$ between target and estimated volatility across test periods for the all the 360 long/short portfolios. Only the best performing window lengths are plotted for each type of approach.}
    \label{fig:All.long.short}
\end{figure*}
\begin{figure*}[!htbp]\centering
    \includegraphics[width=0.8\linewidth]{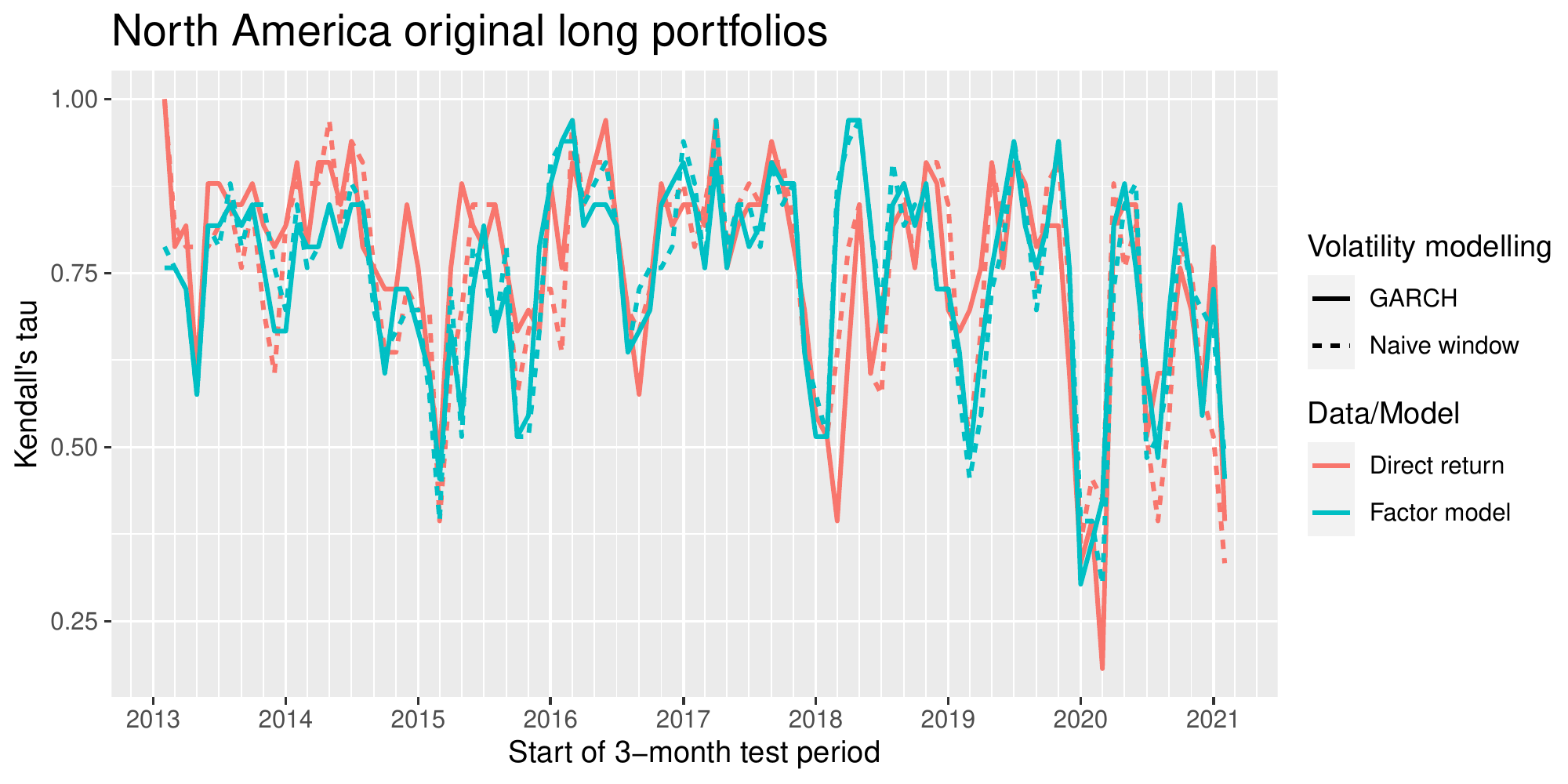}
    \caption{Kendall's $\tau$ between target and estimated volatility across test periods for the 12 original North American style based portfolios. Only the best performing window lengths are plotted for each type of approach.}
    \label{fig:North.America.original.long}
\end{figure*}

Similar results are also obtained for the subset of all the 360 long/short portfolios, see Figure \ref{fig:All.long.short} and Table \ref{tab:All.long.short}. 

While the factor model is not able to rank these large sets of portfolios as well as the direct return approach, it does perform comparably and sometimes better for other, smaller subsets of portfolios. Figure \ref{fig:North.America.original.long} and Table \ref{tab:North.America.original.long} show corresponding results for 12 style based portfolios restricted to North America.
While Table \ref{tab:North.America.original.long} shows the factor model is not as good as the direct approach overall, the performance differences are not big. 
From Figure \ref{fig:North.America.original.long}, we see that at least starting from 2016, the performance of the factor model is generally as good as that of the direct approaches. Further, in the first half of 2018, it is performing significantly better.  

\begin{table}[ht]
\centering
\begin{tabular}{lllr}
  \toprule
Data/Model & Volatility modelling & q & avg($\tau$) \\ 
  \midrule
Direct return & GARCH & 6 & 0.761 \\ 
  Direct return & GARCH & 3 & 0.756 \\ 
  Direct return & Naive window & 6 & 0.753 \\ 
  Direct return & GARCH & 12 & 0.751 \\ 
  Direct return & Naive window & 3 & 0.747 \\ 
  Factor model & GARCH & 6 & 0.745 \\ 
  Factor model & Naive window & 6 & 0.745 \\ 
  Direct return & GARCH & 1 & 0.730 \\ 
  Factor model & GARCH & 12 & 0.729 \\ 
  Factor model & GARCH & 3 & 0.728 \\ 
  Direct return & Naive window & 12 & 0.725 \\ 
  Factor model & Naive window & 12 & 0.722 \\ 
  Factor model & Naive window & 3 & 0.718 \\ 
  Direct return & Naive window & 1 & 0.712 \\ 
  Factor model & GARCH & 1 & 0.694 \\ 
  Factor model & Naive window & 1 & 0.684 \\ 
   \bottomrule
\end{tabular}
\caption{Average Kendall tau for portfolio group 'North America original long'. q denotes the window length in months for the correlation structure (and also the variance structure for the naive window method)} 
\label{tab:North.America.original.long}
\end{table}


In addition to the three portfolio subsets we have looked at above, we have performed analysis for 12 additional portfolio subsets.
The tables with average $\tau$ values for these are provided in Appendix \ref{app:tables}. 
The results for subsets containing only original portfolios and only random portfolios which otherwise are of the same type, are largely similar. 
This indicates that the results are not very sensitive to the companies included in the original portfolios and their specific weights.


\section{Conclusion}
\label{sec:summary}

In this note we have described a set of volatility estimation methods, and studied their ability to rank portfolios according to their future volatility. This included both a direct return approach and using Exabel's factor model, combined with either a GARCH or naive window based volatility modelling approach. 
In the estimation universe considered here, we have seen that the direct return approach is the overall best method. 
However, for some portfolios the factor model has similar or even slightly better performance over some time periods.
It is also clear from the analysis that a GARCH model is almost always preferable over a naive window based volatility modelling approach. The best performing window length $q$ used for the correlation structure (and also the variance for the naive window based approach), seems to vary quite a bit between portfolio subsets. This is expected as the fluctuations in the volatility might be very different in different portfolio types. For most of the portfolio subsets considered here, window lengths of $q=6$ or 12 months are the best performing ones.



\phantomsection
\bibliographystyle{unsrt}
\bibliography{bibliography.bib}

\begin{thebibliography}{1}

\bibitem{kendall1938new}
M.~G. Kendall.
\newblock A new measure of rank correlation.
\newblock {\em Biometrika}, 30(1/2):81--93, 1938.

\bibitem{Grotmol-exabels-factor-model}
Ø. Grotmol, M.~Scheuerer, K.~Aas, and M.~Jullum.
\newblock Exabel’s factor model, 2021.

\bibitem{bollerslev1986generalized}
T.~Bollerslev.
\newblock Generalized autoregressive conditional heteroskedasticity.
\newblock {\em Journal of econometrics}, 31(3):307--327, 1986.

\bibitem{higham2002computing}
N.~J. Higham.
\newblock Computing the nearest correlation matrix—a problem from finance.
\newblock {\em IMA journal of Numerical Analysis}, 22(3):329--343, 2002.

\bibitem{agresti2010analysis}
A.~Agresti.
\newblock {\em Analysis of ordinal categorical data}, volume 656.
\newblock John Wiley \& Sons, 2010.

\end{thebibliography}


\newpage
\appendix

\section{Additional result tables for portfolio subsets}
\label{app:tables}

In Tables \ref{tab:All.original.long}- \ref{tab:North.America.random.long}, we provide result with average $\tau$ values for 12 additional portfolio subsets. The subsets (with number of portfolios in parenthesis) are:
"All original long" (166), "All original long/short" (120), "All random long" (332), "All random long/short" (240), "Unrestricted original long" (51), "Unrestricted original long/short" (11), 
"Unrestricted random long" (102), 
"Unrestricted random long/short" (22), 
"North America original long" (12),
"North America original long/short" (11), "North America random long" (24), "North America random long/short" (22).

\newpage
\hspace{1cm}
\vspace{0.8cm}
\begin{table}[ht]
\centering
\begin{tabular}{lllr}
  \toprule
Data/Model & Volatility modelling & q & avg($\tau$) \\ 
  \midrule
Direct return & GARCH & 12 & 0.587 \\ 
  Direct return & GARCH & 6 & 0.580 \\ 
  Direct return & GARCH & 3 & 0.568 \\ 
  Direct return & Naive window & 6 & 0.561 \\ 
  Direct return & Naive window & 3 & 0.548 \\ 
  Direct return & Naive window & 12 & 0.546 \\ 
  Direct return & GARCH & 1 & 0.532 \\ 
  Direct return & Naive window & 1 & 0.505 \\ 
  Factor model & GARCH & 6 & 0.471 \\ 
  Factor model & Naive window & 6 & 0.471 \\ 
  Factor model & GARCH & 12 & 0.467 \\ 
  Factor model & GARCH & 3 & 0.464 \\ 
  Factor model & Naive window & 12 & 0.461 \\ 
  Factor model & Naive window & 3 & 0.456 \\ 
  Factor model & GARCH & 1 & 0.435 \\ 
  Factor model & Naive window & 1 & 0.423 \\ 
   \bottomrule
\end{tabular}
\caption{Average Kendall tau for portfolio group 'All original long'.} 
\label{tab:All.original.long}
\end{table}

\hspace{1cm}
\vspace{2.5cm}
\begin{table}[ht]
\centering
\begin{tabular}{lllr}
  \toprule
Data/Model & Volatility modelling & q & avg($\tau$) \\ 
  \midrule
Direct return & GARCH & 12 & 0.667 \\ 
  Direct return & GARCH & 6 & 0.665 \\ 
  Direct return & Naive window & 6 & 0.657 \\ 
  Direct return & GARCH & 3 & 0.649 \\ 
  Direct return & Naive window & 12 & 0.642 \\ 
  Direct return & Naive window & 3 & 0.638 \\ 
  Direct return & GARCH & 1 & 0.615 \\ 
  Direct return & Naive window & 1 & 0.601 \\ 
  Factor model & Naive window & 6 & 0.538 \\ 
  Factor model & Naive window & 3 & 0.535 \\ 
  Factor model & GARCH & 6 & 0.533 \\ 
  Factor model & GARCH & 3 & 0.530 \\ 
  Factor model & Naive window & 12 & 0.524 \\ 
  Factor model & GARCH & 12 & 0.522 \\ 
  Factor model & GARCH & 1 & 0.516 \\ 
  Factor model & Naive window & 1 & 0.512 \\ 
   \bottomrule
\end{tabular}
\caption{Average Kendall tau for portfolio group 'All original long/short'.} 
\label{tab:All.original.long.short}
\end{table}

\begin{table}[ht]
\centering
\begin{tabular}{lllr}
  \toprule
Data/Model & Volatility modelling & q & avg($\tau$) \\ 
  \midrule
Direct return & GARCH & 12 & 0.570 \\ 
  Direct return & GARCH & 6 & 0.564 \\ 
  Direct return & GARCH & 3 & 0.552 \\ 
  Direct return & Naive window & 6 & 0.550 \\ 
  Direct return & Naive window & 3 & 0.534 \\ 
  Direct return & Naive window & 12 & 0.534 \\ 
  Direct return & GARCH & 1 & 0.511 \\ 
  Direct return & Naive window & 1 & 0.485 \\ 
  Factor model & GARCH & 6 & 0.468 \\ 
  Factor model & Naive window & 6 & 0.468 \\ 
  Factor model & GARCH & 12 & 0.461 \\ 
  Factor model & GARCH & 3 & 0.461 \\ 
  Factor model & Naive window & 12 & 0.455 \\ 
  Factor model & Naive window & 3 & 0.453 \\ 
  Factor model & GARCH & 1 & 0.432 \\ 
  Factor model & Naive window & 1 & 0.419 \\ 
   \bottomrule
\end{tabular}
\caption{Average Kendall tau for portfolio group 'All random long'.} 
\label{tab:All.random.long}
\end{table}

\begin{table}[ht]
\centering
\begin{tabular}{lllr}
  \toprule
Data/Model & Volatility modelling & q & avg($\tau$) \\ 
  \midrule
Direct return & Naive window & 6 & 0.591 \\ 
  Direct return & GARCH & 6 & 0.590 \\ 
  Direct return & GARCH & 12 & 0.590 \\ 
  Direct return & Naive window & 12 & 0.575 \\ 
  Direct return & GARCH & 3 & 0.573 \\ 
  Direct return & Naive window & 3 & 0.569 \\ 
  Direct return & GARCH & 1 & 0.524 \\ 
  Factor model & Naive window & 6 & 0.522 \\ 
  Factor model & GARCH & 6 & 0.521 \\ 
  Factor model & Naive window & 3 & 0.518 \\ 
  Factor model & GARCH & 3 & 0.516 \\ 
  Factor model & GARCH & 12 & 0.511 \\ 
  Direct return & Naive window & 1 & 0.509 \\ 
  Factor model & Naive window & 12 & 0.505 \\ 
  Factor model & GARCH & 1 & 0.498 \\ 
  Factor model & Naive window & 1 & 0.494 \\ 
   \bottomrule
\end{tabular}
\caption{Average Kendall tau for portfolio group 'All random long/short'.} 
\label{tab:All.random.long.short}
\end{table}

\begin{table}[ht]
\centering
\begin{tabular}{lllr}
  \toprule
Data/Model & Volatility modelling & q & avg($\tau$) \\ 
  \midrule
Direct return & GARCH & 12 & 0.575 \\ 
  Direct return & GARCH & 6 & 0.569 \\ 
  Direct return & GARCH & 3 & 0.558 \\ 
  Direct return & Naive window & 6 & 0.546 \\ 
  Direct return & Naive window & 3 & 0.538 \\ 
  Direct return & Naive window & 12 & 0.530 \\ 
  Direct return & GARCH & 1 & 0.512 \\ 
  Direct return & Naive window & 1 & 0.484 \\ 
  Factor model & Naive window & 6 & 0.447 \\ 
  Factor model & Naive window & 3 & 0.440 \\ 
  Factor model & GARCH & 6 & 0.439 \\ 
  Factor model & GARCH & 3 & 0.437 \\ 
  Factor model & GARCH & 12 & 0.433 \\ 
  Factor model & Naive window & 12 & 0.431 \\ 
  Factor model & GARCH & 1 & 0.414 \\ 
  Factor model & Naive window & 1 & 0.404 \\ 
   \bottomrule
\end{tabular}
\caption{Average Kendall tau for portfolio group 'Unrestricted original long'.} 
\label{tab:Unrestricted.original.long}
\end{table}

\begin{table}[ht]
\centering
\begin{tabular}{lllr}
  \toprule
Data/Model & Volatility modelling & q & avg($\tau$) \\ 
  \midrule
Direct return & GARCH & 12 & 0.692 \\ 
  Direct return & Naive window & 12 & 0.676 \\ 
  Direct return & GARCH & 6 & 0.676 \\ 
  Direct return & Naive window & 6 & 0.671 \\ 
  Direct return & GARCH & 3 & 0.655 \\ 
  Direct return & Naive window & 3 & 0.649 \\ 
  Direct return & GARCH & 1 & 0.636 \\ 
  Direct return & Naive window & 1 & 0.626 \\ 
  Factor model & Naive window & 6 & 0.622 \\ 
  Factor model & GARCH & 12 & 0.616 \\ 
  Factor model & GARCH & 6 & 0.612 \\ 
  Factor model & Naive window & 12 & 0.610 \\ 
  Factor model & GARCH & 3 & 0.594 \\ 
  Factor model & Naive window & 3 & 0.593 \\ 
  Factor model & GARCH & 1 & 0.564 \\ 
  Factor model & Naive window & 1 & 0.551 \\ 
   \bottomrule
\end{tabular}
\caption{Average Kendall tau for portfolio group 'Unrestricted original long/short'.} 
\label{tab:Unrestricted.original.long.short}
\end{table}

\begin{table}[ht]
\centering
\begin{tabular}{lllr}
  \toprule
Data/Model & Volatility modelling & q & avg($\tau$) \\ 
  \midrule
Direct return & GARCH & 12 & 0.551 \\ 
  Direct return & GARCH & 6 & 0.547 \\ 
  Direct return & GARCH & 3 & 0.535 \\ 
  Direct return & Naive window & 6 & 0.529 \\ 
  Direct return & Naive window & 3 & 0.517 \\ 
  Direct return & Naive window & 12 & 0.515 \\ 
  Direct return & GARCH & 1 & 0.485 \\ 
  Direct return & Naive window & 1 & 0.461 \\ 
  Factor model & Naive window & 6 & 0.436 \\ 
  Factor model & GARCH & 6 & 0.431 \\ 
  Factor model & Naive window & 3 & 0.430 \\ 
  Factor model & GARCH & 3 & 0.428 \\ 
  Factor model & GARCH & 12 & 0.422 \\ 
  Factor model & Naive window & 12 & 0.422 \\ 
  Factor model & GARCH & 1 & 0.406 \\ 
  Factor model & Naive window & 1 & 0.399 \\ 
   \bottomrule
\end{tabular}
\caption{Average Kendall tau for portfolio group 'Unrestricted random long'.} 
\label{tab:Unrestricted.random.long}
\end{table}

\begin{table}[ht]
\centering
\begin{tabular}{lllr}
  \toprule
Data/Model & Volatility modelling & q & avg($\tau$) \\ 
  \midrule
Direct return & Naive window & 6 & 0.592 \\ 
  Direct return & GARCH & 6 & 0.579 \\ 
  Direct return & GARCH & 12 & 0.579 \\ 
  Direct return & Naive window & 12 & 0.575 \\ 
  Direct return & GARCH & 3 & 0.556 \\ 
  Direct return & Naive window & 3 & 0.552 \\ 
  Factor model & GARCH & 6 & 0.543 \\ 
  Factor model & Naive window & 6 & 0.541 \\ 
  Factor model & GARCH & 12 & 0.534 \\ 
  Factor model & Naive window & 12 & 0.530 \\ 
  Factor model & Naive window & 3 & 0.524 \\ 
  Factor model & GARCH & 3 & 0.524 \\ 
  Direct return & GARCH & 1 & 0.492 \\ 
  Factor model & GARCH & 1 & 0.484 \\ 
  Direct return & Naive window & 1 & 0.480 \\ 
  Factor model & Naive window & 1 & 0.475 \\ 
   \bottomrule
\end{tabular}
\caption{Average Kendall tau for portfolio group 'Unrestricted random long/short'.} 
\label{tab:Unrestricted.random.long.short}
\end{table}

\begin{table}[ht]
\centering
\begin{tabular}{lllr}
  \toprule
Data/Model & Volatility modelling & q & avg($\tau$) \\ 
  \midrule
Direct return & GARCH & 6 & 0.761 \\ 
  Direct return & GARCH & 3 & 0.756 \\ 
  Direct return & Naive window & 6 & 0.753 \\ 
  Direct return & GARCH & 12 & 0.751 \\ 
  Direct return & Naive window & 3 & 0.747 \\ 
  Factor model & GARCH & 6 & 0.745 \\ 
  Factor model & Naive window & 6 & 0.745 \\ 
  Direct return & GARCH & 1 & 0.730 \\ 
  Factor model & GARCH & 12 & 0.729 \\ 
  Factor model & GARCH & 3 & 0.728 \\ 
  Direct return & Naive window & 12 & 0.725 \\ 
  Factor model & Naive window & 12 & 0.722 \\ 
  Factor model & Naive window & 3 & 0.718 \\ 
  Direct return & Naive window & 1 & 0.712 \\ 
  Factor model & GARCH & 1 & 0.694 \\ 
  Factor model & Naive window & 1 & 0.684 \\ 
   \bottomrule
\end{tabular}
\caption{Average Kendall tau for portfolio group 'North America original long'.} 
\label{tab:North.America.original.long.app}
\end{table}

\begin{table}[ht]
\centering
\begin{tabular}{lllr}
  \toprule
Data/Model & Volatility modelling & q & avg($\tau$) \\ 
  \midrule
Direct return & Naive window & 6 & 0.732 \\ 
  Direct return & GARCH & 6 & 0.731 \\ 
  Direct return & GARCH & 12 & 0.730 \\ 
  Direct return & Naive window & 3 & 0.728 \\ 
  Direct return & GARCH & 3 & 0.726 \\ 
  Direct return & Naive window & 12 & 0.722 \\ 
  Direct return & Naive window & 1 & 0.697 \\ 
  Direct return & GARCH & 1 & 0.696 \\ 
  Factor model & Naive window & 6 & 0.637 \\ 
  Factor model & Naive window & 12 & 0.633 \\ 
  Factor model & GARCH & 6 & 0.623 \\ 
  Factor model & Naive window & 3 & 0.622 \\ 
  Factor model & GARCH & 3 & 0.609 \\ 
  Factor model & GARCH & 12 & 0.606 \\ 
  Factor model & GARCH & 1 & 0.598 \\ 
  Factor model & Naive window & 1 & 0.590 \\ 
   \bottomrule
\end{tabular}
\caption{Average Kendall tau for portfolio group 'North America original long/short'.} 
\label{tab:North.America.original.long.short}
\end{table}

\begin{table}[ht]
\centering
\begin{tabular}{lllr}
  \toprule
Data/Model & Volatility modelling & q & avg($\tau$) \\ 
  \midrule
Direct return & Naive window & 3 & 0.707 \\ 
  Direct return & Naive window & 6 & 0.706 \\ 
  Direct return & GARCH & 3 & 0.705 \\ 
  Direct return & GARCH & 6 & 0.704 \\ 
  Factor model & GARCH & 6 & 0.702 \\ 
  Factor model & Naive window & 6 & 0.698 \\ 
  Direct return & GARCH & 12 & 0.696 \\ 
  Direct return & Naive window & 12 & 0.692 \\ 
  Factor model & GARCH & 12 & 0.687 \\ 
  Factor model & Naive window & 12 & 0.686 \\ 
  Factor model & GARCH & 3 & 0.680 \\ 
  Factor model & Naive window & 3 & 0.680 \\ 
  Direct return & GARCH & 1 & 0.667 \\ 
  Factor model & Naive window & 1 & 0.656 \\ 
  Factor model & GARCH & 1 & 0.655 \\ 
  Direct return & Naive window & 1 & 0.644 \\ 
   \bottomrule
\end{tabular}
\caption{Average Kendall tau for portfolio group 'North America random long'.} 
\label{tab:North.America.random.long}
\end{table}

\begin{table}[ht]
\centering
\begin{tabular}{lllr}
  \toprule
Data/Model & Volatility modelling & q & avg($\tau$) \\ 
  \midrule
Direct return & Naive window & 6 & 0.653 \\ 
  Direct return & GARCH & 6 & 0.645 \\ 
  Direct return & GARCH & 12 & 0.638 \\ 
  Direct return & Naive window & 12 & 0.638 \\ 
  Direct return & GARCH & 3 & 0.628 \\ 
  Direct return & Naive window & 3 & 0.626 \\ 
  Direct return & GARCH & 1 & 0.586 \\ 
  Direct return & Naive window & 1 & 0.570 \\ 
  Factor model & GARCH & 6 & 0.564 \\ 
  Factor model & Naive window & 6 & 0.564 \\ 
  Factor model & GARCH & 12 & 0.552 \\ 
  Factor model & Naive window & 12 & 0.551 \\ 
  Factor model & Naive window & 3 & 0.546 \\ 
  Factor model & GARCH & 3 & 0.545 \\ 
  Factor model & GARCH & 1 & 0.515 \\ 
  Factor model & Naive window & 1 & 0.510 \\ 
   \bottomrule
\end{tabular}
\caption{Average Kendall tau for portfolio group 'North America random long/short'.} 
\label{tab:North.America.random.long.short}
\end{table}

\end{document}